# PROSPECTS FOR QUANTUM COMPUTING: EXTREMELY DOUBTFUL

M. I. DYAKONOV

*Université Montpellier II, CNRS*
*Montpellier 34095, France*
*Michel.Dyakonov@univ-montp2.fr*

The quantum computer is supposed to process information by applying unitary transformations to $2^N$ complex amplitudes defining the state of $N$ qubits. A useful machine needing $N \sim 10^3$ or more, the number of continuous parameters describing the state of a quantum computer at any given moment is at least $2^{1000} \sim 10^{300}$ which is much greater than the number of protons in the Universe. However, the theorists believe that the feasibility of large-scale quantum computing has been *proved* via the "threshold theorem". Like for any theorem, the proof is based on a number of assumptions considered as axioms. However, in the physical world none of these assumptions can be fulfilled exactly. Any assumption can be only *approached* with some *limited* precision. So, the rather meaningless "error per qubit per gate" threshold must be supplemented by a list of the *precisions* with which *all* assumptions behind the threshold theorem should hold. Such a list still does not exist. The theory also seems to ignore the undesired *free evolution* of the quantum computer caused by the energy differences of quantum states entering any given superposition. Another important point is that the hypothetical quantum computer will be a system of $10^3 - 10^6$ qubits PLUS an extremely complex and monstrously sophisticated classical apparatus. This huge and strongly nonlinear system will generally exhibit instabilities and chaotic behavior.

## 1. Introduction

During almost 20 years one can hardly find an issue of any science digest magazine, or even of a serious physical journal, that does not address quantum computing. Computer scientists are proving and publishing new theorems related to quantum computers at a rate of one article per day. Quantum Information Centers are opening all over the globe,[a] and very soon the happy Kingdom of Bhutan in the Himalayas will be the only country without such a Center. The general public, as well as the funding agencies, have been convinced that quantum computers will be unimaginably powerful compared to the classical computers, and that the quantum technological revolution bringing quantum Internet, quantum security against terrorism, and other quantum miracles is at our doorstep. The word "quantum" has entered households and rock star's vocabulary. High school and University students dream of contributing to this daring enterprise.

---

[a] Note that, so far, no quantum information whatsoever is practically available.

When will we have useful quantum computers? The most optimistic experts say: "In 10 years", others predict 20 to 30 years, and the most cautious ones say: "Not in my lifetime". The present author belongs to the meager minority answering "Not in any foreseeable future", and this paper is devoted to explaining such a point of view.

## 2. The general idea of quantum computing

The idea of quantum computing (QC) is to store information in the values of $2^N$ complex amplitudes describing the wavefunction of $N$ two-level systems (*qubits*), and to process this information by applying unitary transformations (*quantum gates*), that change these amplitudes in a precise and controlled manner. The value of $N$ needed to have a useful machine is estimated to be $10^3$ or more. Thus the number of continuous variables describing the state of a quantum computer at any given moment is at least $2^{1000} \sim 10^{300}$ which is much, much greater than the number of protons in the Universe ($\sim 10^{80}$).

However, the QC theorists have succeeded in transmitting to the media and to the general public the belief that the feasibility of large-scale quantum computing *has been proved* via the famous threshold theorem: once the error per qubit per gate is below a certain value indefinitely long quantum computation becomes feasible, at a cost of substantially increasing the number of qubits needed. Very luckily, the number of qubits increases *only polynomially* with the size of computation, so that the total number of qubits needed must increase from $N = 10^3$ to $N = 10^6 – 10^9$ only.

Thus, theorists claim that the problem of maintaining the fragile quantum amplitudes has been solved in principle: "The theory of fault-tolerant quantum computation establishes that a noisy quantum computer can simulate an ideal quantum computer accurately. In particular, the quantum accuracy threshold theorem asserts that an arbitrarily long quantum computation can be executed reliably, provided that the noise afflicting the computer's hardware is weaker than a certain critical value, the *accuracy threshold*".[1] So now the physicists and engineers need only to work on finding the good candidates for qubits and on approaching the accuracy required by the threshold theorem.

## 3. Experiment versus theory

Experimental studies related to the idea of quantum computing make only a small part of the whole QC literature. They are extremely difficult and represent the very top of the modern experimental technique. The goal of such proof-of-principle experiments is to show the possibility of manipulating small numbers of two-level quantum systems serving as qubits, to realize the basic quantum operations, as well as to demonstrate some elements of quantum algorithms, see Ref. 2 for a brief review. The number of qubits used is below 10, usually from 3 to 5. Apparently, going from 5 qubits to 50 (the goal set by the ARDA Experts Panel 2002/2004 roadmap[3] for the year 2012!) presents hardly surmountable experimental difficulties. Most probably they are related to the simple fact that $2^5 = 32$, while $2^{50} = 1125899906842624$.

By contrast, the theory of quantum computing, which largely dominates in the QC literature, does not appear to meet any substantial difficulties in dealing with millions of qubits. Various noise models are being considered, and it has been proved (under certain assumptions) that errors generated by "local" noise can be corrected by carefully designed and very ingenious methods, involving, among other tricks, *massive parallelism*: many thousands of gates should be applied simultaneously to different pairs of qubits and many thousands of measurements should be done simultaneously too.

### 4.  From axioms to the physical reality

The existing abstract theory of quantum computing is purely mathematical. However, the hypothetical quantum computer is supposed to be a *physical* object, a machine that manipulates very accurately huge numbers of quantum amplitudes, which are continuous complex variables. In mathematics, one proves theorems on the basis of some set of axioms. Thus, to make the necessary link between mathematics and physics and to figure out whether a large-scale quantum computer can be built, it is indispensable to understand whether the axioms used by QC theorists are valid in the physical world.

Unfortunately, as I have explained in detail previously,[2, 4] in the physical world the only strictly valid axiom concerning continuous quantities is my

**Axiom 1.**  *No continuous quantity can have an exact value.*

**Corollary 1.**  *No continuous quantity can be exactly equal to zero.*

Other assumptions (axioms) concerning qubit preparation, gates, noise, measurements, etc, cannot be fulfilled exactly, although they might be *approached* with some *limited* precision.[b] Quantum amplitudes being continuous variables, they can never be *exactly* zero. Thus, in the context of quantum computing, Corollary 1 can be reformulated as:[2, 4]

**Corollary 2.** *Exact quantum states do not exist. Some admixtures of all possible states to any desired state are unavoidable.*

So, the real question is: With what precision should hold *all* the assumptions behind the theorems, "proving" the feasibility of large-scale quantum computing?  So far, there are no clear answers to this crucial question.

### 5.  The technical instruction for fault-tolerant computation

The threshold theorem is in fact a technical instruction for fault-tolerant quantum computation, interspersed with lemmas proving that every step will work as it should. This instruction tells us how, by using imperfect instruments, to control on a microscopic level a million of particles, subject to various sources of decoherence and relaxation. The instruction is overwhelmingly complicated.

---

[b] Many QC theorists hold the view that precision is only a matter of time and resources. Indeed, the number of digits of $\pi$ that one can calculate is a matter of time and computational resources. However, whatever time and resources one has, it will never be possible to measure the length of a stick or the resistance of a wire with a precision $10^{-20}$. One (but not the only) reason is that the very notions of stick length or of wire resistance are approximate. Although there are examples of fantastic precision in physics (the $10^{-12}$ precision of the measured anomalous  magnetic moment of an electron, and the $10^{-14}$  accuracy of the atomic clock) , it is never unlimited.

Below is a quotation from the famous threshold-theorem article by Aharonov and Ben-Or[5] describing a small part of the proposed verification procedure for the so-called "cat state": $2^{-1/2}(|0000000> + |1111111>$.[c]

This procedure itself is only a minor detail in the whole scheme providing the anticipated fault-tolerance.

"We now want to compute whether the bits in the cat state are all equal. For this, note that checking whether two bits are equal, and writing the result on an extra blank qubit, can be easily done by a small circuit which uses Toffoli and NOT gates. We denote this circuit by $S$, and also denote the qubits in the cat state by 1, ...$l$. We add $l - 1$ extra blank qubits, and apply the circuit $S$ first from each even pair of qubits (e.g. the pair of qubits (1, 2), (3, 4)...), to one of the blank qubits; Then apply $S$ from each odd pair of qubits (e.g. the pairs (2, 3), (4, 5)...) to one of the remaining blank qubits. We get $l - 1$ qubits which are all 1, if no error occurred, indicating that all the qubits are equal. We then apply a classical circuit on all these qubits, which checks whether they are all 1, and write the result on an extra blank qubit, which is our check bit, and indicates that all the bits in the cat state are equal. We now want to use the cat state, but condition all the operations on the fact that the check bit is indeed 1. However, if we do this, an error in the check bit, might propagate to all qubits in the state we are trying to correct. Hence, to keep the procedure fault tolerant, we construct $m$ different check bits, one for each qubit in the state we are correcting. This is done using $m(l - 1)$ blank qubits, and applying all the operations above, where each operation is repeated $m$ times to $m$ different target qubits. Thus, we can verify that the cat state is of the form $c_0|0> + c_1|1>$, fault tolerantly. We can now condition all the operations done in the syndrome measurement involving the $i$'th qubit, on the $i$'th check bit."

Leaving the reader to ponder on this text, we note that in a technical instruction one does not say: "Make an Hadamard transformation on $N$ qubits", this does not make any sense for the engineer. Instead, one should say: "Rotate each of your 1000 spins around the y-axis (defined with a precision $\varepsilon_1$) by 90 (plus or minus $\varepsilon_2$) degrees. Initially, the spin should be directed along $z$ with a precision $\varepsilon_3$. Also, take care that the disturbance of other spins by this action is less than $\varepsilon_4$, and that the spin-spin interaction in dimensionless units is less than $\varepsilon_5$." The hypothetical Future Quantum Engineer (FQE) will need *this* kind of instruction together with the values of all epsilons to understand whether QC is feasible.

## 6.  The undesired free evolution of the quantum computer

In theory, so long as there is no noise and no gates are applied, an arbitrary superposition of $2^N$ states of $N$ qubits remains intact, implying that any such superposition is a *stationary state*. This is true only under the condition that the energies of all states in our superposition are *exactly equal*. Otherwise, there will be a free evolution of the system determined by its energy spectrum. This was already pointed out 12 years ago in Ref. 6.

---

[c] The cat state is a tool for encoding the logical qubits by a greater number of physical qubits, which is the cornerstone of the fault-tolerance schemes. There exist other methods for encoding without using the cat state.

For example, if the energy difference between the states |1> and |0> of a single qubit is $\Delta E=\hbar\Omega$, like for a spin in magnetic field, and at $t=0$ we create the state $2^{-1/2}(|0> + |1>)$, then at later times the state will evolve as $2^{-1/2}(|0> + \exp(-i\Omega t)|1>)$, describing the spin precession in the plane perpendicular to magnetic field with frequency $\Omega$. For a more general case, consider at $t=0$ the state, which is the starting point for the Shor's factoring algorithm:

$$\Psi_0 = 2^{-N/2}\sum_x \big| x \big\rangle, \qquad\qquad (1)$$

where $\big| x \big\rangle$ is a state of $N$ qubits corresponding to an integer $x$ from 0 to $2^N - 1$ written in binary. This is not a stationary state and its evolution at $t > 0$ is described by:

$$\Psi(t) = 2^{-N/2}\sum_x A_x(t)\big| x \big\rangle, \qquad A_x(t) = exp(-in_x\Omega t). \qquad (2)$$

Here $n_x$ is the number of qubits in the |1> state within a state $\big| x \big\rangle$, $0 \le n_x \le N-1$. As a result, the relative phases will have random values at the moments when subsequent gates are applied, and this will disorganize the performance of the quantum computer.[c]

One might hope to find a solution of this problem by moving to the frame rotating with the frequency $\Omega$. In the rotating frame the qubit energy levels become degenerate, and there will be no evolution in time. However, this means that the *phases* of ac fields (realizing the gates) acting *at any time* during the computation *on all qubits* should be perfectly matched. In other words the *timing* of all gates should be fixed with a precision much better then $(N\Omega)^{-1}$, which is hardly possible.[d]

In fact, things are still worse, because the values of $\Omega$ for different qubits cannot possibly be *exactly* equal (in spectroscopy, this is described as *inhomogeneous broadening*), so that the spectrum of our system will not simply consist of all harmonics of $\Omega$ from 1 to $N-1$. Rather, it will have a quasi-continuous character resulting in a (quasi-) chaotic behaviour of the wavefunction $\Psi(t)$.

This undesired free evolution consisting in a free regular rotation of all qubits can hardly be described as "noise", and it is not obvious at all how the "future quantum engineer" is supposed to deal with it.

It appears that the only way to avoid this unwanted evolution of our *N*-qubit wavefunction is to use qubits with degenerate states. In this case, there is no physically meaningful way to define |0> and |1> states. However, *exact* degeneracy is never possible (unless it is a consequence of some exact symmetry). There is a multitude of reasons for lifting the degeneracy and splitting the |0> and |1> energy levels. Now, the trick of moving to the rotating plane will not work, so there always will be some chaotic evolution of the system. Besides, since the degeneracy is lifted by some random fields, that are generally different for different qubits, the meaning of |0> and |1> for one qubit is not the same as for another one. To perform gates, we will have to use quasi-static magnetic fields to independently rotate individual spins in the desired direction with the desired precision, a truly Sisyphean task.

---

[c] Quite astonishingly, the *energies* of |0> and |1> states and the elementary quantum-mechanical property described above are never a subject of discussion in the abstract QC theory.

[d] Consider, for example, in more detail the creation of the state in Eq. 1 by applying $\pi/2$ microwave pulses to a thousand of spins at different locations, all initially in the |0> states. All the spins should rotate *exactly* in the *xy* plane with *identical* phases. Supposing that we have succeeded, consider the requirements for phase matching needed to perform at a later time the inverse operation of returning all spins back to the state |0>.

In any case, it seems that the QC theoretical vocabulary, consisting of "qubits, gates, noise, and measurements", should be extended to include also the term "energy".[e]

## 7. An optimistic look into the future

Let us assume a highly optimistic scenario that in 10, or 20−30, or 1000 years, the technology will advance to such a perfection that all technical and other problems are resolved and we are finally able to realize the dream of factoring numbers ~$10^{200}$ by Shor's algorithm. The Future Quantum Engineer (FQE) will then ask for a *technical instruction*: what should be done, in what order, and with what precision. At present, it is not clear at all, how and when such an instruction can appear.

Assuming that it will be created some day, it should be noted that the FQE will not be able to check the state and adjust the functioning of his quantum computer because this requires knowing the values of (at least) $10^{300}$ quantum amplitudes, which is impossible. So, he will have to just follow his instructions, construct the hardware, introduce the number to be factored as input, run the machine, and hope for the best: that the final measurement of 1000 logical qubits[f] will give the correct answer.

What if it does not? How is the FQE supposed to find the bug(s)?

A similar problem was put forward by Levin[7]:

"QC proponents often say they win either way, by making a working QC or by finding a correction to Quantum Mechanics. … Consider, however, this scenario. With few q-bits, QC is eventually made to work. The progress stops, though, long before QC factoring starts competing with pencils. The QC people then demand some noble prize for the correction to the Quantum Mechanics. But the committee wants more specifics than simply a nonworking machine, so something like observing the state of the QC is needed. Then they find the Universe too small for observing individual states of the needed dimensions and accuracy. (Raising sufficient funds to compete with pencil factoring may justify a Nobel Prize in Economics.) … So, what thought experiments can probe the QC to be in the state described with the accuracy needed? I would allow to use the resources of the entire Universe, but not more! "

If the instruction for factoring one-hundred-digit numbers is too difficult to develop, maybe somebody could produce a technical manual for factoring 15 with error correction? (Full Shor's algorithm, please, no "compiled" versions).

---

[e] In specific numerous proposals for quantum computing with spins in quantum dots, trapped ions, etc., the energies of quantum states are obviously introduced as an important issue. However, these proposals mostly concentrate on initialization, implementing gates, and read-out, and do not, to my knowledge, discuss the consequences of the free evolution addressed in this Section.

[f] This implies measuring the state of $10^6$ to $10^9$ physical qubits, depending on the level of concatenation.

In the case that even this simple task does not arouse any interest in the QC community, what about quantum computing the identity 1=1, meaning that we just want to use the methods of error correction to store a given state of $N$ qubits for a time exceeding the decoherence time, say, by a factor of 100 or 1000 (quantum memory).

Let us start by $N$=1, a challenge that I posed in Ref. 8. Storing just one qubit is obviously the simplest meaningful problem in the field of fault-tolerant quantum computing. Remember, we are *not* talking about an experimental demonstration, which might be achieved not so soon, if ever. It is a purely theoretical question, and its interest lies in the fact that since the number of qubits involved is not very large, it might be possible to simulate the process on a laptop.

Presumably, to maintain our single qubit close to its initial state $a|0> + b|1>$ with given $a$ and $b$, a certain sequence of operations (with possible branching depending on the result of intermediate measurements) should be applied periodically.

*Provide a full list of these elementary operations*, so that anybody can use his PC to check whether qubit storage really works under the condition that both gates and measurements are imperfect, and what degrees of imperfections and noise are acceptable. *If* it works, this demonstration would be a convincing, though partial, proof that the idea of fault-tolerant quantum computation is sound. So far, this challenge was never met.[g]

## 8. The physical quantum computer as a non-linear system

The *abstract* universal quantum computer, as first defined by Deutsch[9], is an assembly of qubits, with which certain well defined operations are feasible. However, since these operations will not fall from the sky, the eventual *physical* quantum computer will be an assembly of qubits PLUS a monstrously complex and sophisticated classical apparatus, needed to efficiently control many thousands or maybe millions of qubits.[h]

Now, Quantum Mechanics is a linear theory and when a unitary gate $U$ acts on the state $a|0> + b|1>$, one gets $aU|0> + bU|1>$. However measurements are *not* linear operations. Measurement-based error correction transforms the state $a|0> + b|1>$ to $|0>$ or $|1>$, depending on what term is considered to represent an error, and this is not a linear operation. The required equipment by itself is not a linear device either. Thus the whole machine is a huge and strongly nonlinear system, which generally will exhibit instabilities and chaotic behavior.

The performance of a physical quantum computer will remind snowboarding or surfing. The surfer instinctively continuously measures several parameters and adjusts his muscles to act accordingly. Similarly, the quantum computer is supposed to continuously perform a lot of measurements and to choose further actions depending on their result.

---

[g] The aversion of QC theorists to demonstrating how their methods are supposed to work in the simplest situations is reminiscent of the magician episode described by Mark Twain [10]. The magician could easily tell what the Emperor of the East was doing, but could not guess what the Yankee was doing with his right hand. If you don't know how to protect from errors just one qubit, how can you talk about scalable fault-tolerant quantum computation with millions of qubits?

[h] It is sufficient to visit a lab and take a look at the experimental setup used for controlling 3 qubits.

Taking into account the huge number of control parameters within a quantum computer (much, much greater than what the surfer needs), it is not clear whether stable large-scale quantum computing is achievable, and if yes, under what conditions.

This is an important, but totally unexplored issue. Meanwhile there exists a vast field of applied mathematics dealing with instabilities and control of nonlinear systems, see for example Ref. 11. This literature might be of some interest to QC theorists.

## 9. Conclusions

The problems discussed in this article, as well as some others [2, 4, 8] not mentioned here, raise serious doubts about the future of quantum computing. The tremendous gap between the rudimentary, but very hard, experiments and the extremely developed, but rather careless, QC theory is not likely to be closed anytime soon. Besides, it is still not exactly clear what would be the benefits of eventual quantum computing, and whether they are worth the effort of generations of researchers and engineers.

The unprecedented level of hype and of unfounded promises accompanying the QC enterprise is not a good sign either, as well as the multitude of mostly quite irresponsible proposals of "quantum computing with…". [i]   It is a real pity that they never contain, as they should, the famous Landauer's disclaimer: [12]

*"This scheme, like all other schemes for quantum computation, relies on speculative technology, does not in its current form take into account all possible sources of noise, unreliability and manufacturing error, and probably will not work."*

In summary, the perspectives of quantum computing appear to be extremely doubtful. The conventional argument that in the past a similar skepticism was expressed concerning most of the previous human inventions (flying machines, digital computers, and so on) is not a valid one, because there are also even more examples of ideas that did not and never will work. Skepticism is a normal and healthy attitude in science, as opposed to religion, and it is for the believer to give a convincing proof that the anticipated miracle is about to happen.

---

[i] To appreciate the scale of this activity, the reader is invited to google "quantum computing with".